\documentclass[review]{elsarticle}

\usepackage{amssymb}

\usepackage{amsthm}

\usepackage{amsmath}

\usepackage[a4paper,bindingoffset=0.2in,left=1in,right=1in,top=1in,bottom=1in,footskip=.25in]{geometry}

\usepackage{color,soul}

\usepackage{booktabs}

\usepackage{rotating}

\usepackage[version=4]{mhchem}

\usepackage{caption}
\makeatletter
\def\ps@pprintTitle{%
 \let\@oddhead\@empty
 \let\@evenhead\@empty
 \def\@oddfoot{}%
 \let\@evenfoot\@oddfoot}
\makeatother

\journal{Journal of Membrane Science}

\begin{document}
\biboptions{sort&compress}

\abovedisplayshortskip=-13pt
\belowdisplayshortskip=4pt
\abovedisplayskip=-13pt
\belowdisplayskip=4pt

\begin{frontmatter}



\title{Nernst-Planck transport theory for (reverse) electrodialysis: \\ III. Optimal membrane thickness for enhanced process performance\tnoteref{label1}}

\tnotetext[label1]{\textbf{Please cite this article as:} M. Tedesco, H. V. M. Hamelers, P. M. Biesheuvel, \textit{Nernst-Planck transport theory for (reverse) electrodialysis: III. Optimal membrane thickness for enhanced process performance}, Journal of Membrane Science 565 (2018) 480--487. \\}

\author{M. Tedesco}
\author{H.V.M. Hamelers}
\author{and P.M. Biesheuvel \corref{cor1}} \cortext[cor1]{\emph{Corresponding Author} (P. M. Biesheuvel): maarten.biesheuvel@wetsus.nl} 

\address{Wetsus, European Centre of Excellence for Sustainable Water Technology \\ Oostergoweg 9, 8911 MA Leeuwarden, The Netherlands}

\begin{abstract}
The effect of the thickness of ion exchange membranes has been investigated for electrodialysis (ED) and reverse electrodialysis (RED), both experimentally and through theoretical modeling. By developing a two-dimensional model based on Nernst-Planck theory, we theoretically find that reducing the membrane thickness benefits process performance only until a certain value, below which performance drops. For ED, an optimum thickness can be identified in the range of 10--20 $\mu$m, while for RED the maximum power density is found for membranes that are three times as thick. Model calculations compare well with experimental data collected with a series of homogeneous membranes with the same chemical composition and a thickness in the range of 10--75 $\mu$m. Our results show that the classical picture that membranes should be as thin as possible (as long as they remain pinhole-free and structurally stable) is insufficient, and must be replaced by a more accurate theoretical framework.
\vspace{.3cm} 
\end{abstract}


\end{frontmatter}


\section{Introduction}
\label{sec:Introduction}

\indent Ion exchange membranes (IEMs) are nowadays used in a variety of applications, such as for desalination, electrochemical synthesis, selective ion removal, and energy production~\citep{ran2017}. Aside from well-established technologies such as electrodialysis (ED), diffusion dialysis, and fuel cells, novel processes include reverse electrodialyis (RED)~\citep{yip2016,Tedesco2017b}, shock-ED~\citep{mani2009,schlumpberger2015}, membrane capacitive deionization~\citep{Suss2015}, and hybrid (R)ED-reverse osmosis processes \citep{amy2016}, which contribute to broaden the range of applications for ion exchange membranes.

\indent In all of these technologies, the properties of IEMs play a crucial role in determining process performance. For instance, for ED importance properties of the membranes are electrical resistance and permselectivity (or counterion transport number) \citep{sata2004book}. For other applications, water uptake, hydraulic permeability, swelling degree, permselectivity towards specific ions, and pH- and thermal stability, can be equally relevant.

\indent In contrast to the aforementioned properties, the thickness of IEMs is usually not considered to be of primary importance, compared to other membrane properties such as electrical resistance or ion selectivity. Typical values for the thickness of IEMs are in the range of $\delta_\text{m}$=80--250 $\mu$m \citep{ran2017,nagarale2006}, though membranes as thin as $\sim$10 $\mu$m have recently been commercialized. In general, it is commonly accepted in membrane science that membranes should be made as thin as possible, as long as they remain pinhole-free and mechanically stable. Ideally, reducing the thickness of homogeneous membranes gives a linear reduction of the electrical resistance, hopefully without affecting other key properties of the membrane (e.g., ion exchange capacity, permselectivity, or hydraulic permeability). For instance, in reverse electrodialysis (RED), where reducing the internal stack resistance is crucial to enhance the energy production, a significant increase of the (gross) power production in RED has been observed by reducing the membrane thickness ~\citep{Tedesco2015,Moreno2016}.

\begin{figure}[ht]
\centering
\includegraphics[width=\textwidth]{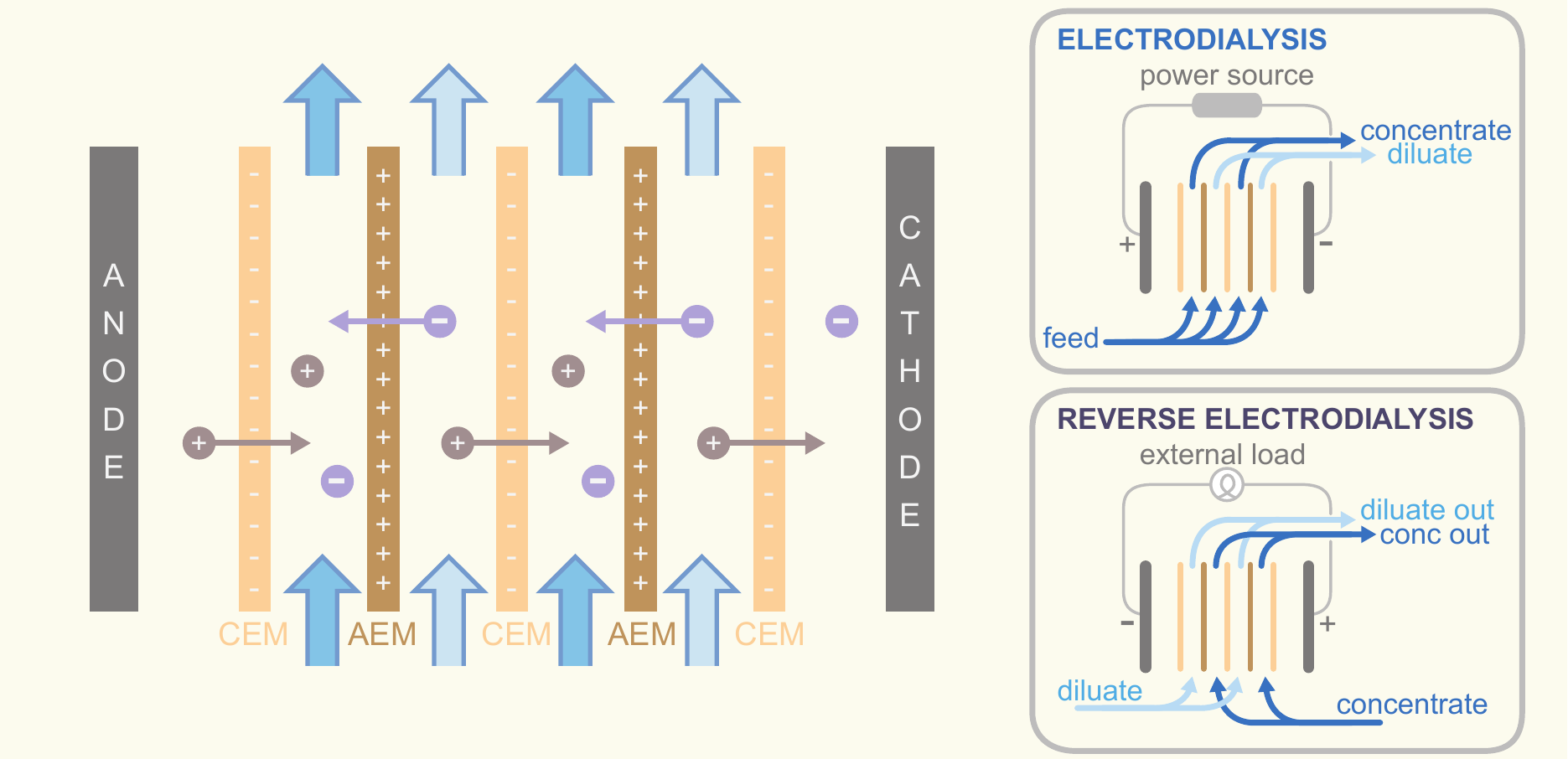}
\caption{Principle of electrodialysis (ED) and reverse electrodialysis (RED).}
\label{fig:FIG_principle}
\end{figure}

\indent According to this common understanding, ``ultra-thin" IEMs (i.e., with $\delta_\text{m}<10$ $\mu$m) should have a performance $\sim$10 times higher than of most of the IEMs currently available in the market. However, 
predicting the actual behavior of such ultra-thin membranes is difficult, since reducing the thickness also affects the transport properties of the membrane (i.e., mass transfer rates for ions, and water permeability). As a consequence, the performance of the membrane can be significantly affected. So far, only a few experimental studies have been reported that describe the effect of membrane thickness on process performance, e.g., for \ce{CO2} gas separation \citep{matsuyama1999}, fuel cells \citep{adachi2010, liu2006, oh2014}, and vanadium redox flow batteries \citep{chen2013}. However, for ED and RED a theoretical framework that describes the behavior of ultra-thin membranes is still missing.

\indent The aim of this work is to investigate the effect of the thickness of ion exchange membranes in ED and RED (Fig.~\ref{fig:FIG_principle}), in order to describe the process when very thin membranes are used. As we will outline further on, by using a model based on the Sonin-Probstein approach \citep{Sonin1968}, implementing the Nernst-Planck equation \citep{Tedesco2016}, we theoretically discover that an optimum value for the membrane thickness can be identified, dependent on details of the process. To validate the model, experiments are performed using a series of homogeneous IEMs with the same chemical composition, and with a variable thickness, in the range of $\delta_\text{m}=10$--75 $\mu$m. By combining this theoretical and experimental information, we aim to develop an understanding of the relation between IEM thickness and other relevant membrane properties, in order to identify the optimal range for the membrane thickness, for a certain process and membrane material.

\section{Theory of ion transport in ion exchange membranes}
\label{sec:theory}

\begin{figure}[ht]
\centering
	\includegraphics[width=0.7\textwidth]{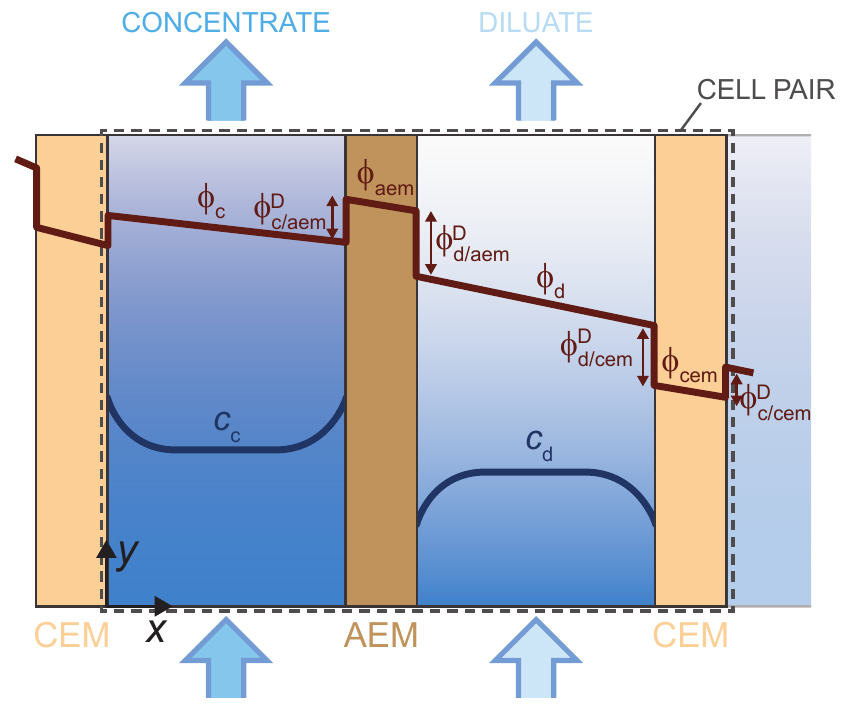}
\caption{Schematic diagram of the model geometry. The computational domain (dashed line) is a full cell pair and consists of one concentrate channel, one anion exchange membrane (AEM), one diluate channel and one cation exchange membrane (CEM). Typical potential and concentration profiles for ED are shown as example.}
\label{fig:FIG_scheme}
\end{figure}

\indent The mass transport theory outlined in this section is based on the two-dimensional model for (R)ED used in refs.~\citep{Tedesco2016,Tedesco2017b}, which is based on the Sonin-Probstein framework \citep{Sonin1968}. In these earlier works, we considered ``symmetrical'' behavior of AEMs and CEMs, i.e., we assumed that the two membranes have the same thickness and transport properties (diffusion coefficient for counter-/coions, and water permeability), and have the same fixed charge density, $X$, except for the difference in sign (i.e., $\omega X>0$ for AEM, and $\omega X<0$ for CEM). The ion diffusion coefficient in the membrane could differ between coion and counterion, but for each ion (e.g., counterion) it was the same in AEM as CEM. In addition, in the spacer channels, the two ions had the same diffusion coefficient. Because of these assumptions, for a 1:1 salt, symmetry could be assumed across the midplane in the diluate and concentrate channel. The modeling domain in refs.~\citep{Tedesco2016,Tedesco2017b} could therefore be simplified to one membrane, half of a diluate channel, and half of a concentrate channel. Instead, in the present work, all of these symmetry-assumptions are dropped, and we include a possible difference in diffusion coefficients between anion and cation both in the membranes and in the spacer channel. In this way, it is possible to include an unequal thickness, porosity and fixed charge density for the two membranes. Since we no longer assume any symmetry in membrane behavior or across a midplane in a spacer channel, the modeling domain in the present work is the whole repeating unit of the system (``cell pair''), including the full width of both diluate and concentrate channels, and also including both membranes (Fig.~\ref{fig:FIG_scheme}).

In the present work, we assume that both streams contain a completely dissociated 1:1 salt (NaCl). The ion mass balance in the flow compartment (spacer channel) is given by

\begin{equation}
	\epsilon \: \frac{\partial{c_i}}{\partial{t}} + \nabla{\cdot J_i} = 0
	\label{eq:mass_balance}
\end{equation}

\noindent where $\epsilon$ is porosity in the spacer channel, $c_i$ is ion concentration, and $t$ is time. The ion flux, $J_{i}$, can be described by the extended Nernst-Planck (NP) equation

\begin{equation}
	J_i=c_i\: \textbf{v} - D_i \: \left(\nabla{c_i} + z_i c_i \nabla{\phi} \right)
	\label{eq:NP}
\end{equation}

\noindent where $\textbf{v}$ is the superficial fluid flow velocity in the channel,  $D_i$ the ion diffusion coefficient, which can be described by $D_i=D_{\infty,i} \cdot \epsilon / \tau $ where $D_{\infty,i}$ is the diffusion coefficient in free solution, and $\tau$ is the tortuosity factor of the spacer. The tortuosity factor is described according to the Bruggeman equation, $\tau=1/\sqrt{\epsilon}$. Note that Eq. (\ref{eq:NP}) does not include dispersion effects, which can be incorporated in Eq. (\ref{eq:NP}) by using a different coefficient for diffusion than for electromigration, as we considered in ref.~\citep{Tedesco2017b}, where dispersion (mixing, leading to flatter concentration profiles) effectively led to $D_\text{diff}>D_\text{migr}$. Instead, in the present work, we use a single value for the diffusion coefficient for a certain ion, as expressed in Eq. (\ref{eq:NP}). This number will have different values in solution and in the two membranes. 

\indent Substitution of Eq. (\ref{eq:NP}) into Eq. (\ref{eq:mass_balance}), and assuming steady--state conditions, $\partial{c_i}/\partial{t}=0$, leads to

\begin{equation}
	D_{i} \: \nabla \cdot {\left(\nabla{c_{i}} + z_{i} c_{i}\nabla{\phi} \right)}  - \nabla \cdot \left({c_{i} \textbf{v}} \right) = 0
	\label{eq:mass_bal_NP}
\end{equation}

\noindent which applies to both ions in the spacer channels. According to the electroneutrality condition, the concentrations of cation and anion at each point in the spacer channels are equal to each other, thus

\begin{equation}
	\sum_{i}{z_{i}c_{i}} = 0
	\label{eq:LEN}
\end{equation}

\noindent and thus $c_+=c_-=c$ at each point in the channels. Adding up Eq. (\ref{eq:mass_bal_NP}) for anions and cations, and taking the difference, results in

\begin{align}
\begin{split}
	\left( {D_{+} + D_{-}}\right) \: \nabla^{2} c +  \left( {D_{+} - D_{-}}\right) \: \nabla \cdot \left( c \nabla \phi \right) - 2 \; \nabla \cdot \left({c\: \textbf{v}} \right) = 0 , \\
		\left( {D_{+} - D_{-}}\right) \: \nabla^{2} c +  \left( {D_{+} + D_{-}}\right) \: \nabla \cdot \left( c \nabla \phi \right) = 0 .
\end{split}
\label{eq:mass_bal_NP2}
\end{align}

\indent These two equations can be simplified when we neglect diffusion and migration in the axial \mbox{($y-$)direction}~\citep{Sonin1968,Tedesco2016}, and by assuming that the fluid velocity profile is already fully developed at the entrance of the channel, and that fluid flow through the membrane can be neglected. Under these conditions, there is no $x-$component to the fluid velocity, and Eq.~(\ref{eq:mass_bal_NP2}) becomes

\begin{align}
\begin{split}
	\left( {D_{+} + D_{-}}\right) \: \frac{\partial ^2 c}{\partial x ^2}  +  \left( {D_{+} - D_{-}}\right) \: \frac{\partial}{\partial x} \left( c \frac{\partial \phi}{\partial x} \right) - 2 \cdot v_\text{y} \frac{\partial c}{\partial y}= 0 , \\	
\left({D_{+} - D_{-}}\right) \: \frac{\partial ^2 c}{\partial x ^2}  +  \left( {D_{+} + D_{-}}\right) \: \frac{\partial}{\partial x} \left( c \frac{\partial \phi}{\partial x} \right) = 0 .
\end{split}
\label{eq:mass_bal_NP4}
\end{align}

\indent For the fluid flow profile, which we assume to be established right at the entrance of the channel, we assume plug flow, i.e., $v_y$ is independent of $x$ ~\citep{Tedesco2016,Tedesco2017b}. 

\indent For the membranes, we assume that diffusion and electromigration only occur in the $x-$direction (just as in the spacer channel, see Fig. \ref{fig:FIG_scheme}), and we solve Eq.~(\ref{eq:mass_bal_NP}) for zero fluid velocity, $\textbf{v}=0$, combining with electroneutrality,

\begin{equation}
	c_+ - c_- + \omega X= 0
	\label{eq:LENmem}
\end{equation}

\noindent where $\omega$ is the sign of the fixed membrane charge ($\omega=+1$ for AEMs, and $\omega=-1$ for CEMs), and $X$ is the molar concentration of membrane charge, defined per unit volume of solution phase in the membrane \cite{Galama2013}.

Evaluated in the $x-$direction only, the balances for cations and anions result in

\begin{equation}
	\frac{\partial ^2 c_\text{T,m}}{\partial ^2 x}  -\omega X  \frac{\partial ^2 \phi}{\partial x ^2} = 0 \: \quad \quad , 
	\quad \quad 
	\frac{\partial}{\partial x} \left( c_\text{T,m} \frac{\partial \phi }{\partial  x} \right) = 0
\label{eq:mass_bal_NP6}
\end{equation}

\noindent where $c_\text{T,m}=c_{+,\text{m}} +c_{-,\text{m}}$ is the total ions concentration in the membrane. Note that Eq.~(\ref{eq:mass_bal_NP6}) is independent of the values of diffusion coefficients, thus it is also valid when $D_{+,\text{m}} \ne D_{-,\text{m}}$. 

The ions flux in the membrane, $J_\text{ions,m}$, and ionic current density (charge flux), $J_\text{ch}$, are calculated as

\begin{equation}
	J_\text{ions,m}=J_{+,\text{m}}+J_{-,\text{m}} \quad \quad ,
	\quad \quad
	J_\text{ch}=J_{+,\text{m}} - J_{-,\text{m}} .
\label{eq:J_ions_m}	
\end{equation}

At each $y-$position in the cell, the current density is independent of $x$, i.e., it has the same value at each position in the membranes and at each position in both of the spacer channels. The ions flux, however, is different everywhere. Only within a given membrane it has a value that is independent of $x$ because our model considers steady-state. Because of this, each ion flux $J_{i,\text{m}}$ from Eq.~(\ref{eq:NP}) (without advective term, i.e., $\textbf{v}=0$), can be integrated from left (``L'') to right (``R'') across a membrane to arrive at

\begin{equation}
J_{i,\text{m}}	= - \frac{ D_{i,\text{m}}}{\delta_\text{m}} \: \left( { c_{i,\text{m,R}}-c_{i,\text{m,L}} }  + z_i  \int_L^R c_{i,\text{m}} \: d \phi\right)
\label{eq:J_ch_mm}
\end{equation}

\noindent where $\delta_\text{m}$ is the membrane thickness, and $D_{i,\text{m}}$ is the ion diffusion coefficient in the membrane, generally $\sim10-100$ lower than in free solution. 

The concentration of an ion, $c_{i,\text{m}}$, at any position in the membrane relates to total ions concentration, $c_\text{T,m}$, and charge density, $X$, according to

\begin{equation}
c_{i,\text{m}}=\frac{1}{2}\left( c_\text{T,m} - z_i \omega X  \right)
\label{eq:c_m}
\end{equation}

\noindent where $z_i=+1$ for the cation, and $z_i=-1$ for the anion. Eqs. (\ref{eq:mass_bal_NP6}--\ref{eq:c_m}) are valid for both the AEM and CEM, and describe the flux of counterions and coions inside each membrane. 

Returning to the spacer channel, here a similar approach can be used for current, but not for ions flux. For current, $J_\text{ch}$, we obtain after integration over a spacer channel

\begin{equation}
J_\text{ch}=  - \frac{1}{\delta_\text{sp}} \left(	\left( {D_{+} - D_{-}}\right)  \: \left({ c_\text{R}-c_\text{L} }\right)  +  \left( {D_{+} + D_{-}}\right)  \int_L^R c_{i} d\phi \right).
\label{eq:J_ch}
\end{equation}

\indent In the spacer channels we can set up an overall mass balance, to relate the average salt concentration (at a certain $y-$position) to the ion fluxes coming through the membranes. For instance, for the concentrate channel, in the geometry of Fig.~\ref{fig:FIG_scheme}, it is

\begin{equation}
-\delta_\text{sp} v_\text{y}\frac{d\langle c \rangle}{dy}+\frac{1}{2}\cdot \left(J_\text{ions,cem}-J_\text{ions,aem}  \right) = 0
\label{eq:overall balance1}
\end{equation}

\noindent where $\langle c \rangle$ denotes an average salt concentration. This balance only needs to be set up for one channel because the total flowrate of salt flowing through the channels is fixed. Thus, we can use an overall mass balance for the cell pair, for instance, for co-current flow, given by

\begin{equation}
	Q_\text{d}  \cdot \left(c_\text{d,in}-\langle c_\text{d} \rangle \right) + Q_\text{c}  \cdot \left(c_\text{c,in}- \langle c_\text{c} \rangle \right) = 0
\label{eq:overall balance2}
\end{equation}

\noindent where $Q_\text{d}$ and $Q_\text{c}$ are the volumetric flow rates through the diluate and concentrate channels (in volume per time). Assuming plug flow in the spacer channel (i.e., $v_\text{y}$ independent of $x$), the average salt concentration in the spacer (evaluated at a certain $y-$position) is given by $\langle c \rangle \delta_\text{sp}=\int_L^R c\: dx$, where $L$ and $R$ refer to the left and right edges of the channel.

\indent We finally must evaluate boundary conditions for gradients in concentration and potential in the spacer channel at the edge with the membranes. In order to numerically solve the model, we need to implement two equations as boundary conditions. In principle, any of the transport equations (based on charge flux, ions flux, or individual ions fluxes, at any of four positions) can be implemented as boundary conditions (and the result of the calculation will be the same in each case). One example is the cation flux from the CEM into the concentrate channel, which is described by

\begin{equation}
J_{+,\text{m}}=-D_{+} \left(\frac{\partial c}{\partial x} + c\: \frac{\partial \phi}{\partial x}  \right)  
\label{eq:overall balance3}
\end{equation}

\noindent where the gradient terms on the right-hand side must be evaluated in the concentrate spacer channel, at the membrane/solution interface.

\indent The final set of equations relates to the electrical  potentials across a cell pair. The cell pair voltage, $\Delta\phi_\text{CP}$, is given by the sum of all the voltage differences across a cell pair in $x-$direction (Fig. \ref{fig:FIG_scheme}), and can be calculated as

\begin{equation}
	\Delta\phi_\text{CP} = \Delta\phi_\text{c} 	
	+ \left( \Delta\phi^\text{D}_\text{c/aem} + \Delta\phi_\text{aem} - \Delta\phi^\text{D}_\text{d/aem} \right)
	+ \Delta\phi_\text{d} 
	+ \left( \Delta\phi^\text{D}_\text{d/cem} + \Delta\phi_\text{cem} - \Delta\phi^\text{D}_\text{c/cem} \right) 	
\label{eq:cp_voltage}
\end{equation}

\noindent where $\Delta\phi_\text{c}$, $\Delta\phi_\text{aem}$, $\Delta\phi_\text{d}$ and $\Delta\phi_\text{cem}$ are the potential drops over the concentrate channel, the anion-exchange membrane, the diluate channel, and the cation-exchange membrane. The remaining four terms on the right-hand side in Eq. (\ref{eq:cp_voltage}) (i.e., $\Delta\phi^\text{D}_\text{c/aem}$, $\Delta\phi^\text{D}_\text{d/aem}$, $\Delta\phi^\text{D}_\text{d/cem}$ and $\Delta\phi^\text{D}_\text{c/cem}$) are related to the Donnan potentials arising at the four solution/membrane interfaces (see Fig. \ref{fig:FIG_scheme}). Note that all terms in Eq. (\ref{eq:cp_voltage}) are dimensionless, and can be multiplied by $R\: T/F$ to obtain a dimensional voltage. For a typical membrane stack with ``undivided'' electrodes, the cell pair voltage, $\Delta\phi_\text{CP}$, is independent of $y-$position, while current density, $J_\text{ch}$, can vary significantly in this direction.

\indent The Donnan potential of a certain solution/membrane interface, $\Delta\phi^\text{D}_\text{s/m}$, is a function of the membrane charge density and of the salt concentration at the interface (on the solution side, i.e., in the spacer) \cite{Helfferich1962,Kontturi2008}. This can be calculated as

\begin{equation}
\omega X =	2\: c^*_\text{s/m}\sinh \left( \Delta\phi^\text{D}_\text{s/m} \right) 
	\label{eq:donnan}
\end{equation}

\noindent where subscript `s' refers to either the concentrate or diluate solution, subscript `m' refers to AEM or CEM, and $c^*_\text{s/m}$ is the salt concentration in the spacer channel at the specific solution/membrane interface. This concentration also determines the total ions concentration in the membrane at the same interface, but on the membrane side, $c_\text{T,s/m}$, according to

\begin{equation}
	c_\text{T,s/m} = 2\: c^*_\text{s/m} \cosh \left(\Delta\phi^\text{D}_\text{s/m} \right) = \sqrt{X^{2} + \left (2\: c^*_\text{s/m} \right )^2} .
	\label{eq:C_Tmem}
\end{equation}

\indent The above equations are discretized in the spatial domain in both the $x-$ and $y-$direction, and solved numerically. The integrals defined in Eqs. (\ref{eq:J_ch_mm},\ref{eq:J_ch}-\ref{eq:overall balance2}) are solved using Simpson's rule.

\section{Experimental}
\label{sec:exp}

\indent A lab-scale stack ($10\times 10$ cm$^2$ electrode area, $N=5$ cell pairs) was used for all measurements. The stack is equipped with Ru-Ir coated Ti electrodes (Magneto Special Anodes, The Netherlands), and woven spacers (spacer thickness $\delta_\text{sp}=200$ $\mu$m, Sefar, Switzerland). A series of homogeneous IEMs (Fumatech, Germany) with the same chemical composition and customized thickness in the range of $\delta_\text{m}=10$--75 $\mu$m was tested (see Table \ref{tab:mem_prop}). All membranes have the same (homogeneous) structure, except for the thickest membranes (Fumasep FAS-PET-75, FKS-PET-75), which contain a non-conductive supporting material. Moreover, the thickest membranes have a slightly lower ion exchange capacity (IEC), as reported by the manufacturer. Water uptake was measured for all the membranes at 20 $^\circ$C in 0.25 M NaCl.

\begin{table}[!ht]
\caption{Physical properties of ion exchange membranes used in this work (Fumasep series, Fumatech GmbH, Germany).}
\label{tab:mem_prop}
	\begin{tabular}{ll|*3c|*2c}
		\toprule
		Type	&	Name		&  \multicolumn{3}{c}{thickness ($\mu$m)}	& IEC$^a$		& 	Water uptake$^b$ \\
				&				& nominal 	& dry$^a$	& wet$^b$			&(meq/g$_{\text{dry polymer}}$)	& 	(g$_{\text{water}}$/g$_{\text{dry polymer}}$)	\\	
		\midrule
		AEM &	FAS-10 			& 10		& 12--17	& 12--17			& 1.6--1.8	& 0.10 \\
			&	FAS-15 			& 15		& 14--17	& 15--19			& 1.6--1.8	& 0.11--0.18 \\
			&	FAS-20 			& 20		& 18--22 	& 20--21			& 1.6--1.8	& 0.11--0.14 \\
			&	FAS-50 			& 50	 	& 42--61	& 47--62			& 1.6--1.8	& 0.23--0.27 \\
			&	FAS-PET-75		& 75		& 77--88	& 78--96			& 1.0--1.4	& 0.18--0.20 \\
		\midrule		
		CEM &	FKS-10 			& 10		& 11--14	& 12--16			& 1.3--1.4	& 0.11--0.16 \\
			&	FKS-15 			& 15		& 11--13	& 12--20			& 1.3--1.4	& 0.06--0.11 \\
			&	FKS-20 			& 20		& 20--26 	& 23--26			& 1.3--1.4	& 0.14--0.19 \\
			&	FKS-55 			& 55	 	& 55--61	& 56--65			& 1.3--1.4	& 0.19--0.21 \\
			&	FKS-PET-75		& 75		& 78--96	& 92--97			& 0.8--1.2	& 0.19 \\

		\bottomrule
	\end{tabular}
	\caption*{$^a$ From manufacturer specifications. $^b$ Measured at 20 $^\circ$C in 0.25 M NaCl.}
\end{table}

\indent An aqueous solution of 0.1 M \ce{K3Fe(CN)6}, 0.1 M \ce{K4Fe(CN)6}, and 0.25 M NaCl was used as electrode rinse solution. The feed solutions were prepared using demineralized water and technical-grade NaCl (Esco, The Netherlands), with a salt concentration of 30 g/L NaCl (i.e., mimicking seawater) for the ED tests, and 30 g/L vs. 1 g/L NaCl in the case of RED tests (i.e. mimicking typical seawater-fresh water conditions). Peristaltic pumps (MasterFlex, Cole-Palmer, USA) were used to pump all solutions (including the electrode rinse solution) into the stack. All experiments were performed at 20 $^\circ$C, with a ``symmetric'' membrane configuration, i.e., with the AEM and CEM having the same thickness. Theoretical calculations reported are also based on $\delta_\text{aem}=\delta_\text{cem}=\delta_\text{m}$, except for the work reported in Fig.~\ref{fig:contour}. 

\section{Results and discussion} 
\label{sec:results}

\subsection{Effect of membrane thickness on current efficiency and energy consumption in ED}
\label{sec:results_ED}

\indent In this section we report both experimental and theoretical results for seawater desalination via electrodialysis, where the experiments are based on a membrane thickness in the range of $\delta_\text{m}=10$--75 $\mu$m. Flow velocity both in the diluate and in the concentrate channels is $v=1$ cm/s (i.e., a residence time of $\tau=10$ s). The spacer channels are modeled assuming a co-current flow arrangement, with a feed concentration of $c_\text{inlet}=500$ mM NaCl, spacer thickness of $\delta_\text{sp}=200$ $\mu$m, $\epsilon/\tau=0.35$, and different ion diffusion coefficients for cations and anions, i.e., $D_{+}=1.33\cdot10^{-9}$ m$^2$/s, and $D_{-}=2.03\cdot10^{-9}$ m$^2$/s. For the membranes, we consider a fixed charge density of $X=4$ M for both AEM and CEM, and an ion diffusion coefficient for all ions in both membranes of $D_{i,\text{m}}=2.5\cdot10^{-11}$ m$^2$/s. The model does not assume a certain electrical resistance for the membranes, but uses only the ion diffusion coefficients, the membrane charge density, $X$, and membrane thickness as input parameters. Therefore, the effect of membrane thickness on process performance can be investigated without any further simplifying assumptions about electrical resistance or permselectivity.

\begin{figure}[!ht]
\centering
\includegraphics[width=0.9\textwidth]{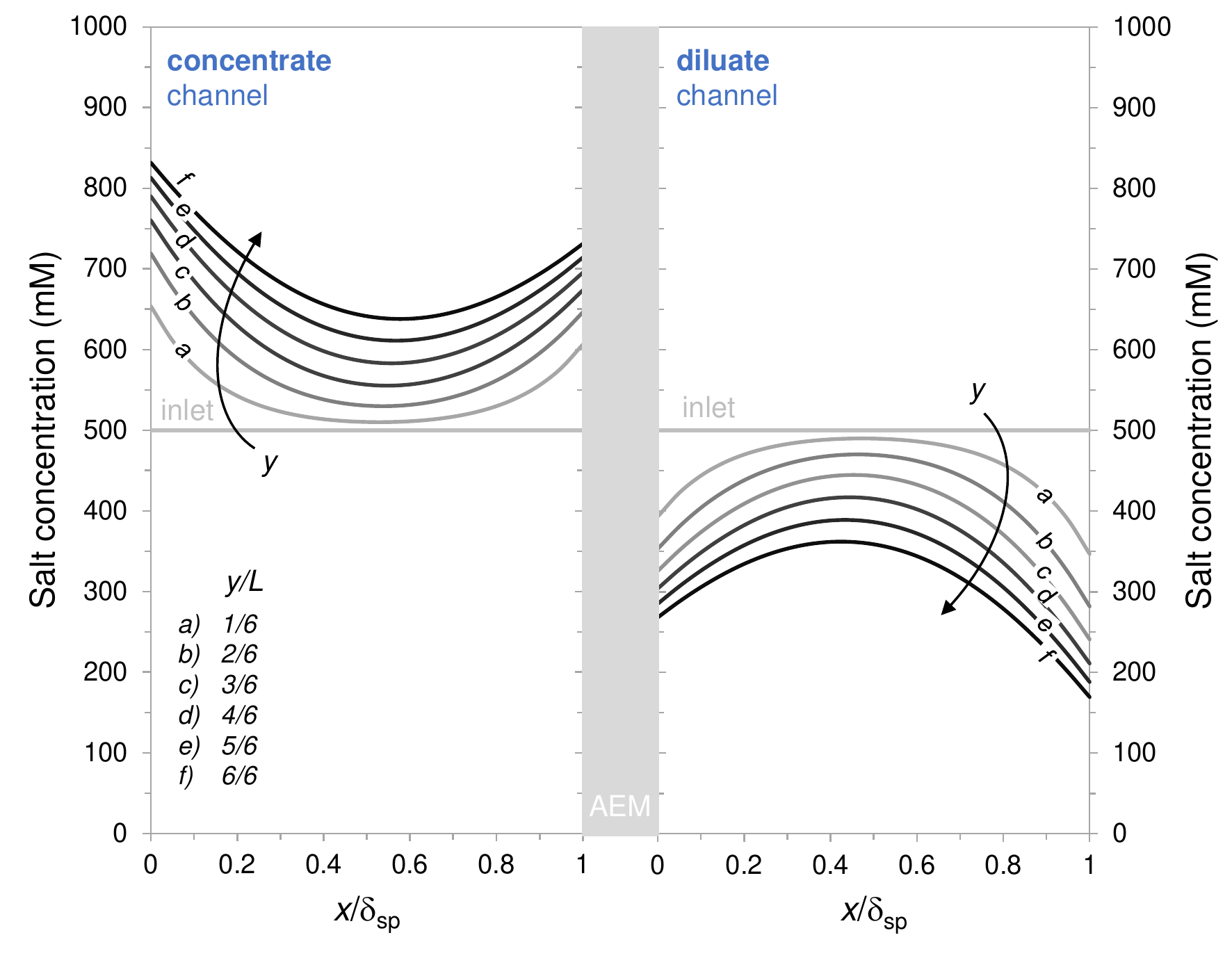}
\caption{Theoretical results for concentration profiles in concentrate and diluate channels in ED as function of $x-$position. Each curve represents a different $y-$position in the channel ($y=0$: inlet; $y=L$: outlet). Current density $I=400$ A/m$^2$.}
\label{fig:ED-conc_prof}
\end{figure}

\indent The model calculates the concentration profiles across the width of the channels, to quantify the effect of polarization phenomena for different membrane properties (i.e., thickness and fixed charge density), and operating conditions (flow velocities and current density). As an example, Fig.~\ref{fig:ED-conc_prof} shows the concentration profiles for both concentrate and diluate channels, where the model was solved with a discretization step of $\Delta y=L/6$, and the concentration profiles at different $y-$position in the channel are shown as a result. Calculation results shown in Fig.~\ref{fig:ED-conc_prof} are based on membranes with a thickness of $\delta_\text{m}=25$ $\mu$m, and for a current density of $I=400$ A/m$^2$. Note that throughout this work, current current density $I$ is calculated as the average current density in the cell thus averaged over the $y-$coordinate, i.e., $I= \frac{F}{L} \:\int_0^L J_\text{ch}\: dy$. We also introduce the average salt flux through the membranes, $\langle J_\text{salt} \rangle$, which is equal to 

\begin{equation}
	\langle J_\text{salt} \rangle = \frac{1}{2\:L} \:\int_0^L \left(J_\text{ions,aem}^*+J_\text{ions,cem}^* \right)\: dy, 
\end{equation}

\noindent where $^*$ is added to signify that the fluxes are defined as positive in the direction out of the diluate channel. For the calculations shown in Fig.~\ref{fig:ED-conc_prof}, this salt flux is $\langle J_\text{salt} \rangle=3.8$ mmol/m$^2$/s. As can be observed, the concentration profiles in Fig.~\ref{fig:ED-conc_prof} are not symmetric, i.e., a different degree of polarization is predicted on the AEM/solution interface than on the CEM/solution interface. We see that polarization is more prominent at the CEM surface, because in solution \ce{Na^+} has a lower diffusion coefficient than \ce{Cl^-}.

\begin{figure}[!ht]
\includegraphics[width=\textwidth]{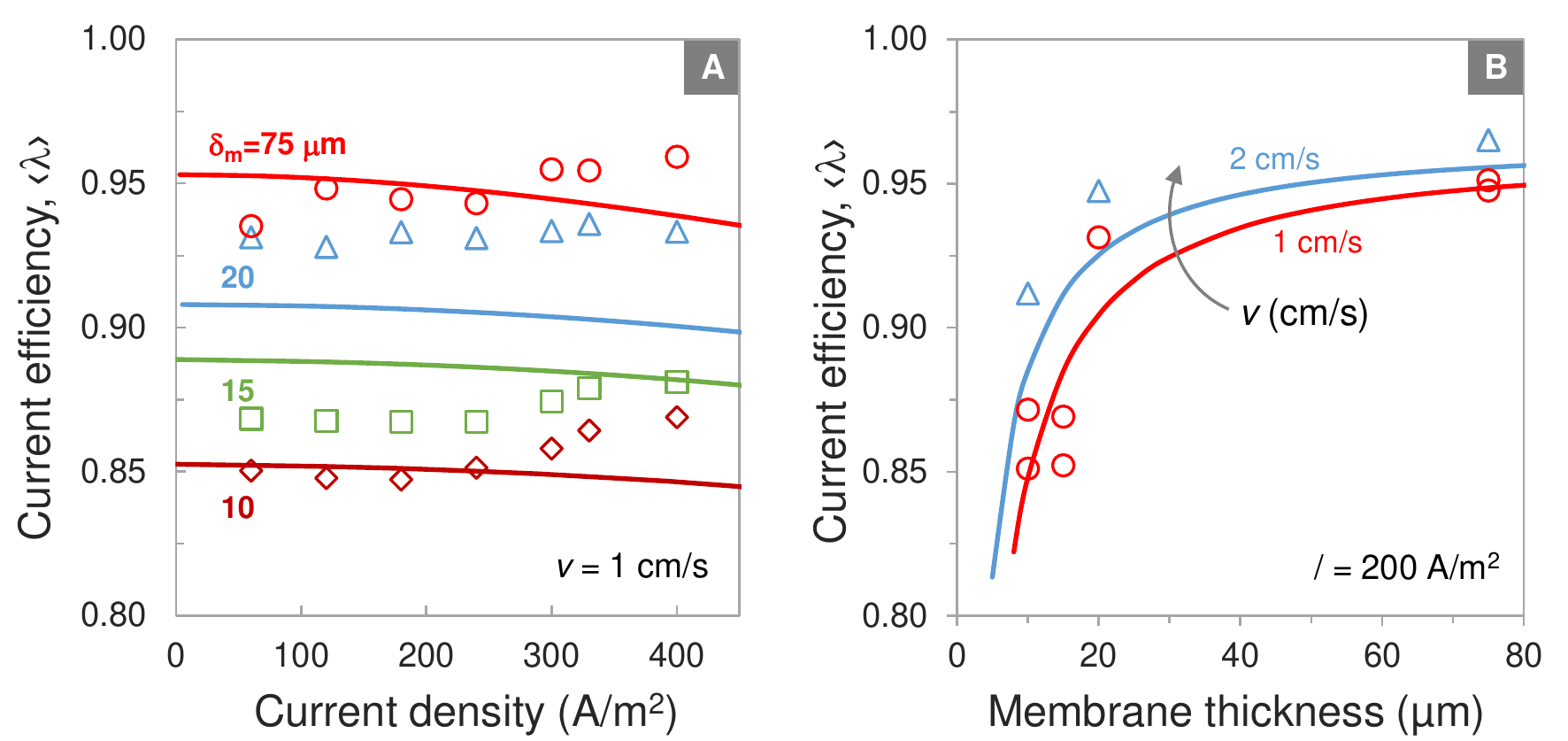}
\caption{ED: Average current efficiency, $\langle \lambda \rangle$, as function of A) current density, $I$, and B) membrane thickness, $\delta_\text{m}$. Comparison between experimental data (symbols) and model predictions (lines) for four different values of the membrane thickness, $\delta_\text{m}$ (AEM and CEM have the same thickness). In B), current efficiency is reported for two different flow velocities, $v$.}
\label{fig:ED-lambda}
\end{figure}

\indent The effect of membrane thickness on the ED process can be quantified by the average current efficiency, $\langle \lambda \rangle$, i.e., the ratio between the average salt flux, $\langle J_\text{salt} \rangle$, and the average ionic current, $\langle J_\text{ch} \rangle=I/F$~\citep{Tedesco2016}. Fig. \ref{fig:ED-lambda} shows results of the current efficiency (both experimentally and theoretically) as function of current density, $I$ (Fig. \ref{fig:ED-lambda}A), and as function of membrane thickness (Fig. \ref{fig:ED-lambda}B). Current efficiency does not depend much on current density (Fig. \ref{fig:ED-lambda}A), while it decreases with membrane thickness, from $\langle \lambda \rangle=0.95$ for the thickest  membranes ($\delta_\text{m}=75$ $\mu$m), to $\langle \lambda \rangle=0.85$ for the thinnest membranes ($\delta_\text{m}=10$ $\mu$m), with a quite good agreement between model predictions and experimental data. The effect of membrane thickness on $\langle \lambda \rangle$ is shown in Fig. \ref{fig:ED-lambda}B, for one current density $I=200$ A/m$^2$ and for two different flow velocities. Current efficiency is relatively constant for membranes with thickness $\delta_\text{m}>30$ $\mu$m, while it decreases with thickness for $\delta_\text{m}<30$ $\mu$m. For membranes thinner than $\delta_\text{m}=20$ $\mu$m, current efficiency, $\langle \lambda \rangle$, drops to below 90\%. Increasing the flow velocity leads to an increase of current efficiency, especially when $\delta_\text{m}<20$ $\mu$m (Fig. \ref{fig:ED-lambda}B).

\begin{figure}[!ht]
\centering
\includegraphics[width=\textwidth]{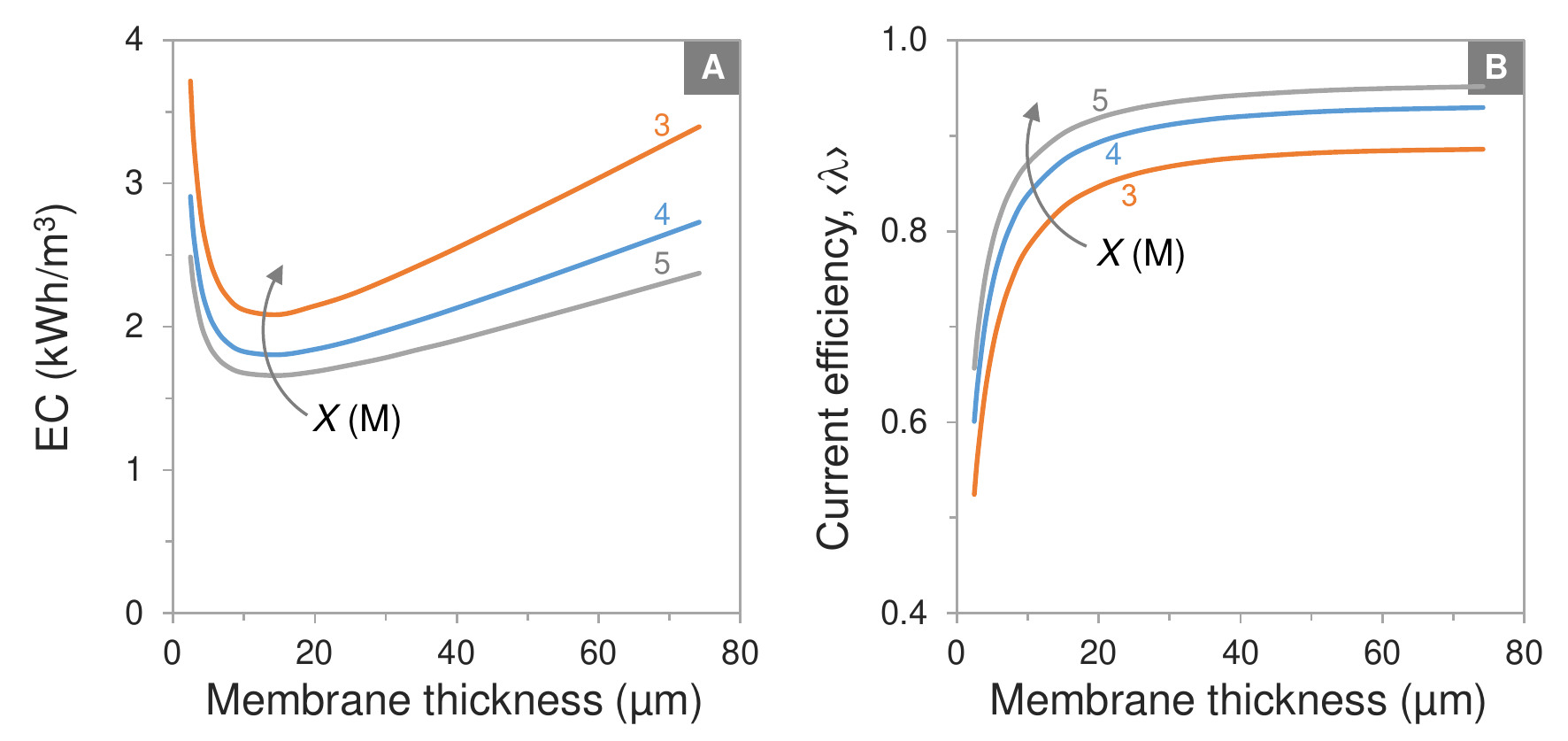}
\caption{ED: Effect of membrane thickness, $\delta_\text{m}$, on A) energy consumption, EC, and B) current efficiency, $\langle \lambda \rangle$, as function of membrane charge density, $X$. Salt flux: $\langle J_\text{salt} \rangle=5$ mmol/m$^2$/s. AEM and CEM have the same thickness, i.e., $\delta_\text{aem}=\delta_\text{cem}=\delta_\text{m}$.}
\label{fig:ED_EC_lambda}
\end{figure}

\indent Next, we investigate the influence of membrane thickness on energy consumption, EC, which is the electrical energy required per volume of diluate produced. In our calculation, we set average salt flux to $\langle J_\text{salt} \rangle=5$ mmol/m$^2$/s. This salt flux, which is a typical experimental value, leads to a 20\% desalination for a flow velocity of $v=1$ cm/s (i.e., the inlet salinity of 500 mM is reduced in the diluate channel to $c_\text{d}=400$ mM). Fig. \ref{fig:ED_EC_lambda} shows the effect of membrane thickness on energy consumption, EC, and on the average current efficiency, $\langle \lambda \rangle$, for different values of membrane charge density, $X$ (in the range of $X$ from 3 M to 5 M). Fig. \ref{fig:ED_EC_lambda}A shows that there is a minimum in the energy consumption when $\delta_\text{m}\sim15$ $\mu$m, and this optimal thickness seems to be independent of $X$. When the membrane thickness is higher than $\delta_\text{m}=15$ $\mu$m, the energy consumption decreases with decreasing thickness, because the electrical resistance of the membranes decreases (approximately linearly with $\delta_\text{m}$ \citep{guler2013performance}). However, the average current efficiency, $\langle \lambda \rangle$, also decreases with decreasing $\delta_\text{m}$ (Fig. \ref{fig:ED_EC_lambda}B), due to the increase of coion leakage. As a result, in the case of membranes with thickness $\delta_\text{m}<15$ $\mu$m, the energy consumption increases again.

\indent In summary, Fig. \ref{fig:ED_EC_lambda} shows that the thickness of membranes influences both the electrical resistance and the co-ion leakage, and therefore an optimum thickness can be found. Interestingly, a similar effect of the thickness of IEMs has also been shown in other processes, e.g., for fuel cells \citep{adachi2010, liu2006, oh2014}, and for proton-blocking IEMs in vanadium redox flow batteries \citep{chen2013}.

\subsection{Effect of membrane thickness on open circuit voltage and power production in RED} 
\label{sec:results_RED}

\indent In this section, we report both experimental and model results for reverse electrodialysis (RED), for the same range of values for the membrane thickness. All experiments and model simulations are for a flow velocity of $v=1$ cm/s, and for inlet concentrations of $c_\text{d}=17$ mM and $c_\text{c}=500$ mM. We investigate the effect of membrane thickness on the open circuit voltage, OCV, and on the maximum power density, PD$_\text{max}$. The voltage across the stack under zero-current condition, OCV, is calculated as the sum of the voltages across all cell pairs under zero current (i.e. $\Delta\phi_\text{stack}=N\: \Delta\phi_\text{CP}$, where $N=5$ is the number of cell pairs in this study), thus neglecting any non-ideal effects due to non-homogeneous flow distribution or parasitic currents in the stack \citep{Veerman2008parasitic}. Note that OCV is calculated for the case that $I$, the current density, averaged over the entire cell, is zero, and thus at one position in the cell the local current may be positive, if compensated by a negative current elsewhere. The maximum power density, PD$_\text{max}$, is the maximum value of power output from the system (determined by varying the external load to a value such that the power generated is at a maximum \citep{Veerman2009a}, divided by the total membrane area (i.e., $2\: N\: A$, where $A=0.01$ m$^2$ is the projected area of one membrane).

\begin{figure}[!ht]
\centering
\includegraphics[width=0.55\textwidth]{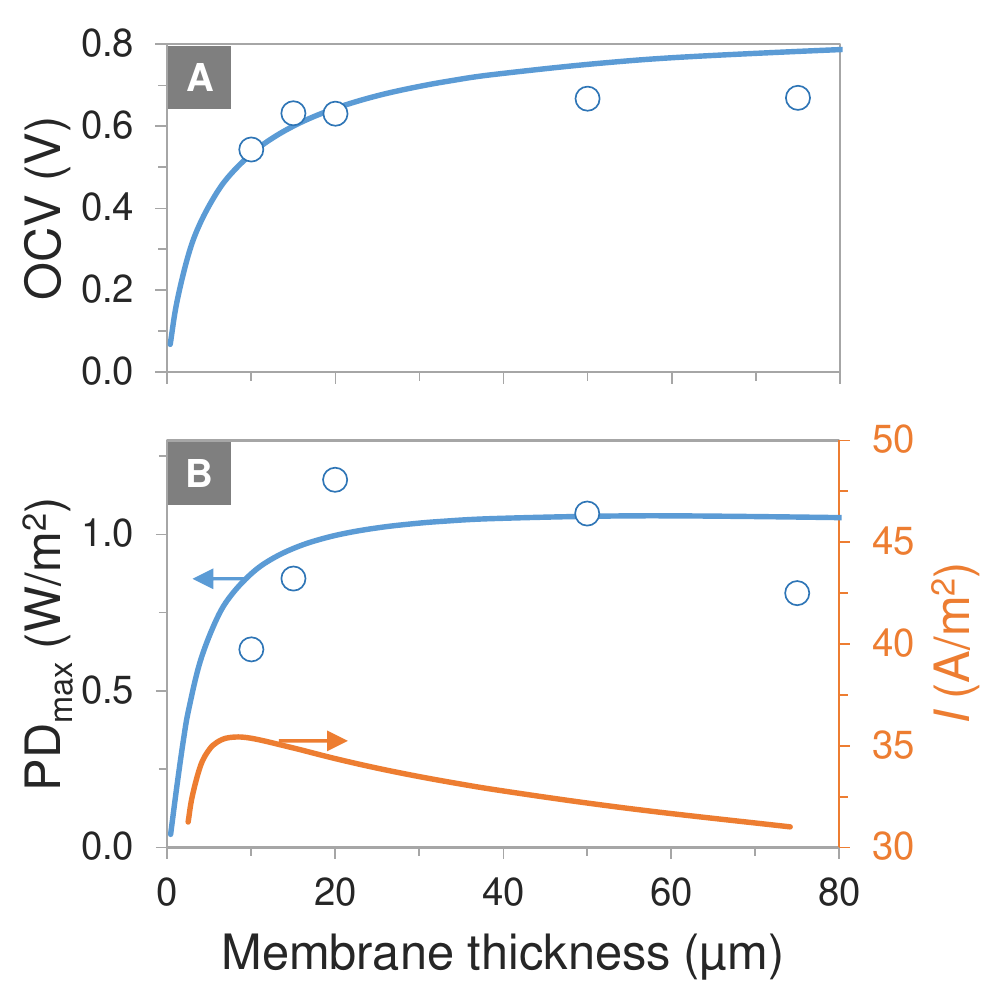}
\caption{RED: Effect of membrane thickness, $\delta_\text{m}$, on A) open circuit voltage, OCV, and on B) maximum power density, PD$_\text{max}$, showing also the corresponding value of current density, $I$. Comparison between experimental data (symbols) and model predictions (lines), using an equal thickness of AEM and CEM ($\delta_\text{aem}=\delta_\text{cem}=\delta_\text{m}$).}
\label{fig:REDvalidation}
\end{figure}

\indent Fig.~\ref{fig:REDvalidation} shows the influence of membrane thickness, $\delta_\text{m}$, on OCV and PD$_\text{max}$, and also shows the current density at PD$_\text{max}$. All theoretical conditions are the same as in Section \ref{sec:results_ED}, except that the diffusion coefficient for all ions in the membrane is increased to $D_{i,\text{m}}=6.5\cdot10^{-11}$ m$^2$/s (which gives a better fitting with the RED data). Calculation results are in good agreement with the experimental data, though the model overestimates the values of PD$_\text{max}$ in the case of $\delta_\text{m}=75$ $\mu$m. This difference is perhaps due to the higher electrical resistance caused by the non-conductive supporting material. According to our theory, membrane thickness does not affect the maximum power density if $\delta_\text{m}>20$ $\mu$m (Fig.~\ref{fig:REDvalidation}). Instead, for membrane thicknesses $\delta_\text{m}<20$ $\mu$m, both OCV and PD$_\text{max}$  decrease significantly when the thickness is reduced, due to the increased coion leakage through thin membranes. The influence of membrane thickness on coion leakage can be inferred from the reduction of the average salt transport efficiency, $\langle \theta \rangle$ (Fig.~\ref{fig:contour}B), because $\langle \theta \rangle$ is inversely proportional to coion leakage \citep{Tedesco2016}.

\subsection{Optimal range of AEM-CEM thickness for ED and RED performance}
\label{sec:results_properties}

\begin{figure}[!ht]
\centering 
\includegraphics[width=.8\textwidth]{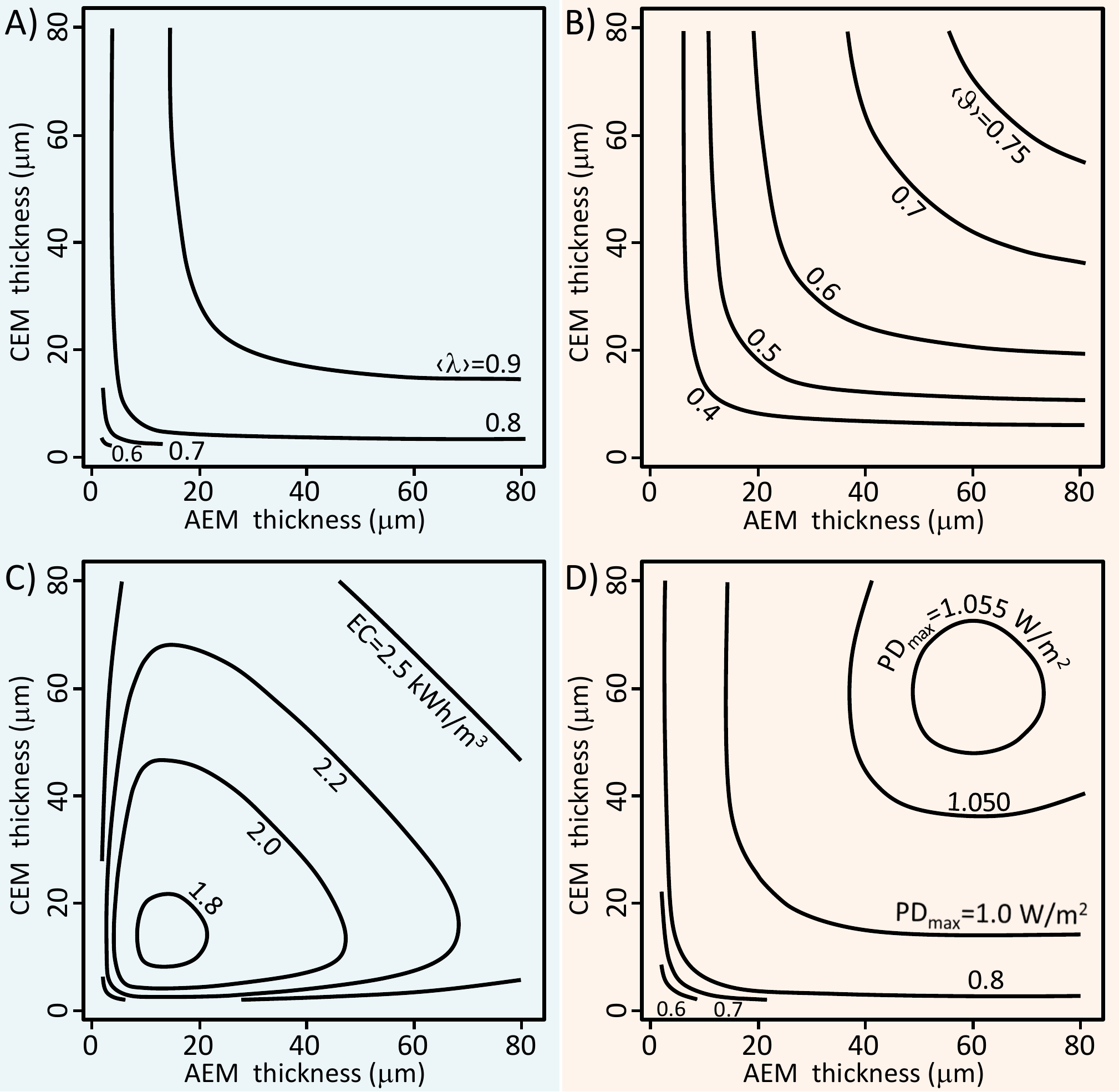}
\caption{Iso-lines shwing the effect of AEM and CEM thickness, $\delta_\text{aem}$, $\delta_\text{cem}$, on A) average current efficiency (ED), $\langle \lambda \rangle$, B) average salt transport efficiency (RED), $\langle \theta \rangle$, C) energy consumption (ED), EC, and D) maximum power density (RED), PD$_\text{max}$. 
}
\label{fig:contour}
\end{figure}

\indent In the previous sections we showed only results for a ``symmetric'' membrane configuration, i.e., where AEM and CEM have the same thickness, $\delta_\text{aem}=\delta_\text{cem}=\delta_\text{m}$. In the present section, we investigate the effect of ``asymmetric'' membrane configurations, i.e., for membrane pairs with different combination of $\left(\delta_\text{aem},\delta_\text{cem} \right)$. In particular, we performed model simulations for all combinations of $\left(\delta_\text{aem},\delta_\text{cem} \right)$, where both $\delta_\text{aem}$ and $\delta_\text{cem}$ vary in the range $\delta_\text{m}=2$--80 $\mu$m. Results are shown in Fig.~\ref{fig:contour}, in terms of average current efficiency, $\langle \lambda \rangle$, and energy consumption, EC, for ED (Fig.~\ref{fig:contour}A,C); and in terms of average salt transport efficiency, $\langle \theta \rangle$, and maximum power density, PD$_\text{max}$, for RED (Fig.~\ref{fig:contour}B,D). All simulations in Fig.~\ref{fig:contour} were performed for a flow velocity of $v=1$ cm/s (i.e., an average residence time $\tau=10$ s), and membrane charge density of $X=4$ M. For ED, we consider a constant average salt flux through the membranes of $\langle J_\text{salt}\rangle=5$ mmol/m$^2$/s. For RED, simulations were performed at the maximum power density. Moreover, we used for these simulations the same ion diffusion coefficient for cations and anions in solution, i.e., $D_\text{i}= \sqrt{D_{+} \cdot D_{-}}=1.64 \cdot 10^{-9}$ m$^2$/s. These assumptions makes the problem symmetric, in the sense that by switching $\delta_\text{aem}$ and $\delta_\text{cem}$ the same results are obtained. Thus, all the plots in Fig.~\ref{fig:contour} are symmetric in the diagonal (i.e., in the line $\delta_\text{aem}=\delta_\text{cem}$). Fig.~\ref{fig:contour} shows ``iso-lines'' that connect pairs of $\left(\delta_\text{aem},\delta_\text{cem} \right)$ giving the same value of the variable considered in each plot.

\indent As can be observed in Fig.~\ref{fig:contour}A, in the case of ED, current efficiency, $\langle \lambda \rangle$, 
weakly depends on membrane thickness, only dropping to below $\langle \lambda \rangle=0.70$ when the membranes become thinner than $\delta_\text{cem}=10$ $\mu$m. However, the effect of membrane thickness on energy consumption is prominent (Fig.~\ref{fig:contour}C), with a minimum value (EC$\sim$1.8 kWh/m$^3$) obtained when both membranes have a thickness in the range of $\delta_\text{m}=15$--20 $\mu$m (Fig.~\ref{fig:contour}C). Note that these calculations, which are based on typical values for salt flux through the membranes and residence time in the channel, correspond to a single-pass ED stage with a low desalination rate ($\sim20$\%). Therefore, the EC values shown in Fig.~\ref{fig:contour}C are applicable to a first stage of seawater desalination process, and the energy consumption to desalinate seawater to drinking water will be much higher. In summary, Fig.~\ref{fig:contour}A,C show that in ED the use of thicker membranes is preferable to reduce coion leakage and enhance $\langle \lambda \rangle$, though this leads to an increase of the electrical resistance and, therefore, of the energy consumption. To minimize the energy consumption, optimal membrane thicknesses are in the range of $\delta_\text{m}=15$--20 $\mu$m (Fig.~\ref{fig:contour}C).

\indent For RED, the average salt transport efficiency, $\langle \theta \rangle$, decreases with decreasing $\delta_\text{m}$, due to the increasing coion leakage through thinner membranes (Fig.~\ref{fig:contour}B). In particular, $\langle \theta \rangle$ ranges from $\langle \theta \rangle=0.40$ (when either $\delta_\text{aem}$ or $\delta_\text{cem}$ is $<15$ $\mu$m), up to $\langle \theta \rangle=0.75$ (with both $\delta_\text{aem}$ and $\delta_\text{cem}$ in the range of 55--80 $\mu$m). Therefore, ``thick'' membranes are preferable to reduce coion leakage, and maximize the salt transport efficiency. However, by increasing $\delta_\text{m}$ also the electrical resistance increases. Therefore, a maximum value of PD$_\text{max}$ is found when both membranes have a thickness of around $\delta_\text{m}=50$--70 $\mu$m (Fig.~\ref{fig:contour}D), which is the result of a trade-off between the reduction in salt transport efficiency (at low $\delta_\text{m}$, Fig.~\ref{fig:contour}B), and the increasing electrical resistance at high $\delta_\text{m}$. Note that for a membrane thickness in the range of $\delta_\text{m}=20$--80 $\mu$m these two effects of reducing selectivity and reducing electrical resistance almost cancel each other, leading to a practically negligible change in PD$_\text{max}$.  

\indent The effect of thickness in Fig.~\ref{fig:contour} is reported for a flow velocity of $v=1$ cm/s for both diluate and concentrate channels. Increasing the flow velocity leads to an increase of this efficiency, especially when $\delta_\text{m}<20$ $\mu$m (see Fig. \ref{fig:ED-lambda}). As a consequence, for flow velocities $v>1$ cm/s the optimal ranges of thickness in Fig.~\ref{fig:contour}C,D is expected to shift towards lower values of $\delta_\text{aem}$ and $\delta_\text{cem}$. However, higher flow velocities lead to higher pressure losses, which can limit the economic feasibility of the process (especially in the case of RED \citep{Tedesco2015}). In conclusion, although operating conditions (i.e., flow velocities and feed concentration) can affect the optimal range of the membrane thickness, we believe that the trends shown in Fig.~\ref{fig:contour} will be generally observed, also for different operating conditions.

\section{Conclusions} 
\label{sec:Conclusions}

\indent The aim of this work has been to investigate the behavior of ion exchange membranes as function of membrane thickness, focusing on thicknesses  below 80 $\mu$m. We showed both theoretically and experimentally that reducing the membrane thickness to $\sim$20 $\mu$m is beneficial for ED, mainly as a result of the reduced electrical resistance. A further reduction of the thickness leads to a decrease of performance because of higher coion leakage. For RED, as long as the membrane thickness is larger than $\delta_\text{m}=20$ $\mu$m, the influence of thickness on selectivity and on electrical resistance almost cancel out, resulting in  a weak dependence of the maximum power density on thickness.

\indent Thus, our work shows that, in the case of membranes thinner than 80 $\mu$m, the effect on process performance of reducing the thickness is not trivial, and a careful analysis is necessary. Ultra-thin membranes, i.e., with thickness $\delta_\text{m}<20$ $\mu$m, can suffer from high coion leakage under actual process conditions. It is important to note that our calculation results are limited to the case of a 1:1 salt solution, while the presence of other ions in solution also affects process performance. Future studies should therefore include the effect of ion mixtures on the performance of ED and RED.

\section*{Acknowledgments}

This work was performed in the cooperation framework of Wetsus, European Centre of Excellence for Sustainable Water Technology (www.wetsus.eu). Wetsus is co--funded by the Dutch Ministry of Economic Affairs and Ministry of Infrastructure and Environment, the Province of Frysl\^{a}n, and the Northern Netherlands Provinces. We thank Fumatech GmbH, Germany, for supplying membranes with customized thickness, and Mr. C. Ersoy for his contribution to the experiments. Furthermore, we thank the participants of the research theme ``Blue Energy'' for fruitful discussions and financial support.

\section*{References}

\bibliographystyle{elsarticle-num} 
\bibliography{Manuscript}


\end{document}